# Exploration of Various Deep Learning Models for Increased Accuracy in Automatic Polyp Detection

Ariel E. Isidro
Computer Engineering Department,
Technological Institute of the
Philippines- Quezon City
Quezon City, Philippines
ariel.isidro@yahoo.com

Arnel C. Fajardo
Computer Engineering Department,
Technological Institute of the
Philippines – Quezon City
Quezon City, Philippines
acfajardo2011@gmail.com

Alexander A. Hernandez
College of Information Technology
Education
Technological Institute of the
Philippines – Manila
City of Manila, Philippines
alexander.hernandez@tip.edu.ph

**ABSTRACT**

This paper is created to explore deep learning models and algorithms that results in highest accuracy in detecting polyp on colonoscopy images. Previous studies implemented deep learning using convolution neural network (CNN) algorithm in detecting polyp and non-polyp. Other studies used dropout, and data augmentation algorithm but mostly not checking the overfitting, thus, include more than four-layer modelss. Rulei Yu et.al from the Institute of Software, Chinese Academy of Sciences said that transfer learning is better talking about performance or improving the previous used algorithm. Most especially in applying the transfer learning in feature extraction. Series of experiments were conducted with only a minimum of 4 CNN layers applying previous used models and identified the model that produce the highest percentage accuracy of 98% among the other models that apply transfer learning. Further studies could use different optimizer to a different CNN modelsto increase accuracy.

**KEYWORDS**
Colon cancer, Overfitting, Dropout, Deep Learning, Transfer Learning

## 1 INTRODUCTION

Colon cancer can be prevented if detected at an early stage since mostly Filipinos are between poor to average ones, it is one of the most fatal cancers in the Philippines[1]. Colonoscopy is the preferred method for screening and preventing colorectal cancer (CRC). CRC usually starts with a growth of a polyp in the inner surface of the colon which can later develop into a cancer[2].

Technology arises machine learning became important application to detect diseases such as CRC. Machine Learning is very useful in creating devices like the detection of colon cancer to images and other object detection[3]. In the previous machine-learning study such as computer vision is based from CADe and/or CADx (Computer Aided Detection and diagnosis)colonoscopy was proposed for system development and clinical testing to detect colonoscopy[4].

The Convolution Neural Network is the most popular computer vision used as image classification in a deep learning algorithm[5].Deep Learning was also used in agricultural area such as leaf disease detection, land cover classification and other fields of training models[6]. Apparently, deep learning used large dataset to increase accuracy[7].This situation has changed with the appearance of the large-scale dataset in deep learning models. One of the great success of CNN is to use large dataset for training[8].A study improve the performance by training a network on all 15 million images and 22,000 ImageNet classes[9]. Mostly deep learning uses Convolutional Neural Network (CNN) in sharing modules and their weights between multiple neural networks to a heterogeneous distributed infrastructure [10]that can be used to improve medical image classification. Simply to apply transfer of learning to different task weight initialization is a part of implementation. The most popular deep learning approach is the supervised pre-training[9] for transfer learning. Later transfer of learning from previous tasks was said to be effective to conduct series of test and improve the performance of the study. As mentioned that pre-trained transfer learning technique is when the learner must perform two or more different tasks[9] in short the datasets are the same while the application to different environment is different.

Conventional CNN architectures make it suitable for training with smaller datasets[7]. Therefore, one of the architectures used in this paper is the deep learning in a simple CNN layer model. Also, small datasets with different models can produce high accuracy. One of the methods of improving the accuracy of the small datasets is using data augmentation algorithm. In the study of classifying fish species produces 99 percent accuracy using the existing CNN model VGG16, a deep convolutional neural network with retraining of the model, fine-tuning, data augmentation and optimization[11]. Likewise VGG pre-trained on 1.2 million images from the ImageNet database[12] shows its good output and VGG19 was used to detect polyp.

This paper also used Adam optimizer like the study that develop deep learning solutions for medical applications[3]. Apparently, researchers used to minimize the overfitting in machine learning algorithm to train the datasets much faster and make better predictions[13], thus, dropout can improve performance[13] because it drops the hidden unit during training[14]. In order to realize transfer learning, application of Adam optimizer, dropout and over-fitting to series of experiments are used to Simple CNN, Augmentation and VGG19 models. The following section discusses related study for polyp detection and deep learning models. Section III is about the method used while the fourth part is the results and discussion.

## 2 RELATED WORKS
This section discusses the related works about polyp detection and the deep learning models.

## 2.1 POLYP DETECTION

Automated polyp detection uses computer programs as an alternative and better solution. The suggested solution examines the images that are obtained during colonoscopy where it determines whether a polyp is present or not. The paper published by Tajbakhsh[15] attained a 96% success rate in detecting polyp. But the technique used here showed tedious manual pre-processing. The automated polyp detection technique consisted of four stages: (1) constructing an edge map for an input image, (2) refining the edge map by classifying every edge pixel into polyp and non-polyp categories using context information, (3) localizing polyp candidates from the refined edge maps using shape information, and (4) placing a band around each polyp candidate to measure the probability of being a polyp.

The use of image processing and analysis in the medical industry posed a great potential in automating polyp detection and reducing loss of life due to colon-cancer. Studies proved that this technology has greatly reduced miss detection of polyps which leads to late detection of colon cancer[15]. However, the success rate registered by above mentioned researches may still require further studies must to be conducted as the highest success rate in polyp detection. After reading previous studies 96% is the highest accuracy[16]and this leads significant amount of complex manual pre-processing.

Besides, automatic colon polyp detection using Convolutional Encoder-Decoder model with Auto encoders, batch normalization, relu, pooling, up sampling, softmax[17] was used. Polyp detection using regression-based CNN focuses to ResYOLO algorithm and came up with 88.6 % precision[18]. Furthermore, the paper about convolutional neural network (CNN) for polyp detection that is constructed based on Single Shot MultiBox Detector (SSD) architecture[19]with max pooling layers has the precision of 90.4%.

## 2.2 DEEP LEARNING MODELS

Deep Learning Models often composes of layers with different operations and functions conducted to come up an output easily especially to a large scale of data sets. Mostly the learning of features is better or the learning is faster, better generalization to new data, and better predictions when using the final models[3] with large parameters. The exploration of deep learning models which is figuring automatically the right set of features or simply features learning[3]from a raw data. Fully connected deep neural networks, Convolutional Neural Networks (CNNs), recurrent neural networks (RNNs), Long Short-Term Memory (LSTM) networks are used. Researchers also mentioned Biomedicine NiftyNet[20] as deep learning platform that supports image segmentation, regression, generation and representation learning applications for medical imaging. Moreover, computer vision was used in deep learning framework that produces better recognition accuracy and higher robustness in complex environment[21]. Latest advances of CNN focuses to many tasks such as image classification, object detection, object tracking, pose estimation, text detection, visual saliency detection, action recognition, scene labelling, speech and natural language processing[8]. The detection of erosions and ulcerations in deep neural network[19]was also applied in deep learning Convolution Neural Network.

Polyp detection using CNN-SVM (Convolution Neural Network-Support Vector Machine)classifier was used in different studies and Alexnet as feature extractor[22]. Additionally, Haibing Wu and XiaodongGu[13]propose to use dropout at the max pooling layer. The said study has different way of pre-trained CNN with proper fine-tuning received a lot of attention in various signal processing fields. It focuses a dictionary learning algorithm and classified polyps in 95% accuracy, sensitivity, specificity, and precision.

## 3 METHODOLOGY

Simple Convolutional Neural Network(CNN) is the trained model. The IPO chart below shows the basic structure of a Convolutional Neural Network. The input image is shown in Fig. 1as the input to CNN process which outputs the classification where it is polyp or non-polyp image.

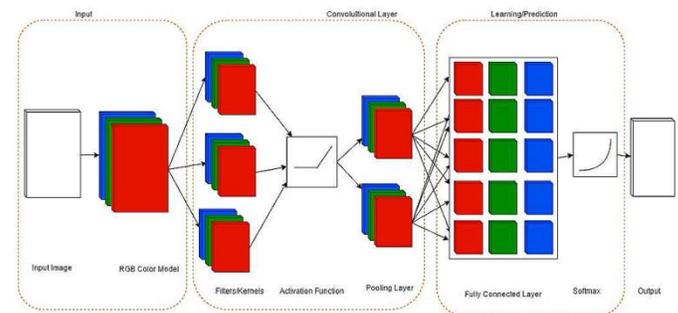

**Figure1. Convolutional Neural Network (CNN) involves three major process: (a) Input Image (b) Convolution (c) Prediction[23]**

Fig.1 illustrates a typical convolutional neural network. It consists of three main elements which are the input image, the convolutional layer and the prediction layer[23]. The input image is the original image to be trained. Given that the input is a colored image, then it shall be decomposed according to the intensities of the red, green, and blue as described in the RGB color model. It is CNN layer normally a three-stage layer starting from the convolution. During this stage, the input image is convolved using a predetermined filter or kernel yield a matrix called a feature map. By convolving, the dot product of the kernel and the region under it is computed. The kernel is moved using a sliding window technique pixel-by-pixel. After convolution, the activation function is triggered. Several activation functions could be used but the most popular one is the Rectified Linear Unit function (ReLU). This function retains only the positive values and removes all the negative values. After activation is the pooling layer. At the pooling layer, several algorithms could also be used such as average pooling, max pooling interprets the output of the model. But the most commonly used algorithm in this paper is the max pooling where the maximum value for each sliding windows is retained leaving all the smaller values.

Medicine opens the database CVC colon DB to all researchers. CVC-ClinicDB is a database used by different authors to increase accuracy percentage of detecting polyp on the image for colon cancer patient. The experiments results conducted uses a database from CVC-ClinicDB[24] and the trained dataset consist of 968 different images. The dataset was divided into two classes with labels polyp and normal. Thereafter, the dataset was split into three sets, 80% for training, 10% for validation and the remaining 10% for testing.

## 3.1 DATA PREPARATION

Data Preparation/Data Pre-processing describe how the data is processed into a usable form such as missing data imputation, typecasting, handling duplicates and outliers[25], and so on. During the data preparation, the polyp and non-polyp (normal) file are first generated from the original dataset of polyp-images and corresponding ground truth. Once the images are generated the image are resized and scaled into usable form since the output created [26]are not uniform in size and width. Fig. 2 and 3 are the sample scaled image.

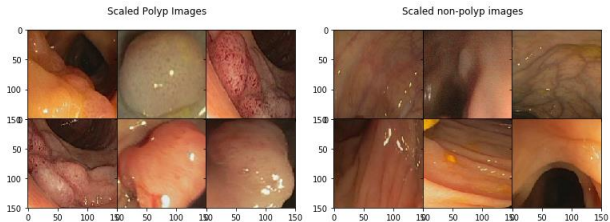

**Figure 2. Scaled Images**

Figure 2 left shows the scaled images with polyp while figure on the right shows scaled normal colon image. The scaled image became small by dividing the images by 255.

Keras provides access to a number of top-performing pre-trained models which is usually for image recognition tasks[25]. Keras was used as neural network library then the layers Convolutional 2D grid of pixels, MaxPooling2D as a feature vector representation, Flatten is for performance checker, and Dense is used to predict the probability of classes. Dropout is added to increase accuracy and the Adam as Optimizer.

## 3.2 MODEL TRAINING

Model Training involves many small linear algebra operations, such as matrix multiplications and dot products[3]to identify the results of the used model. This study uses supervised because the training image data was tallied to the image label.

There are 968 trained datasets to different model. A data generator serves as input to the deep learning model during model training. The generated polyp and non-polyp (normal) images were joined in one training file and identify its length. This came up with single database. The dataset was loaded so scale was used before training it to different models. Also, high epoch was used to check which values result in high accuracy rate. Models used are 1) Simple CNN, 2) Augmentation and 3) VGG19 as Feature Extractor with different value of dropout. The three model were used because data augmentation model reduces over-fitting[11], VGG19 is a popular transfer learning model while a simple CNN has three to four convolution neural network layers.

Max pooling, early stopping, model checkpoint, and patience is equivalent to 200. Early stopping and model checkpoint were used to stop the training after the maximum validation accuracy is reached while the maximum training epoch is 3000.

**Table 1 Model Name and its description**

| Model | Description | Dropout | Epoch | Elapsed (min) |
|---|---|---|---|---|
| Simple CNN | | | | |
| M1-1 | 3CNN | | 208 | 3.677 |
| M1-1-Best | 3CNN | | 15 | |
| M1-2 | 4CNN | | 210 | 2.59762 |
| M1-2-Best | 4CNN | | 32 | |
| M1-3 | 4CNN | 0.3 0.3 | 213 | 2.72607 |
| M1-3-Best | 4CNN | 0.3 0.3 | 191 | |
| M1-4 | 4CNN | 0.3 0.3 | 255 | 3.18475 |
| M1-4-Best | 4CNN | 0.3 0.3 | 113 | |
| M1-5 | 4CNN | 0.5 0.5 | 247 | 3.1145 |
| M1-5-Best | 4CNN | 0.5 0.5 | 33 | |
| Augmentation | | | | |
| M2-1 | Augment | | 741 | 77.78 |
| M2-Best | Augment | | 401 | |
| M2-2 | Augment | 0.3 0.3 | 575 | 56.5516 |
| M2-2-Best | Augment | 0.3 0.3 | 375 | |
| M2-3 | Augment | 0.5 0.5 | 695 | 72.74423 |
| M2-3-Best | Augment | 0.5 0.5 | 332 | |
| VGG19 Feature Extractor | | | | |
| M3-4 | | 0.3 0.3 | 318 | 37.232 |
| M3-4-Best | | 0.3 0.3 | 76 | |
| M3-5 | Augment | | 728 | 80.41008 |
| M3-5-Best | Augment | | 440 | |
| M3-6 | Augment | 0.5 0.5 | 588 | 58.8122 |
| M3-6-Best | Augment | 0.5 0.5 | 311 | |
| M3-7 | Augment | | 662 | 65.27 |
| M3-7-Best | Augment | | 233 | |
| M3-8 | Fine Tune Augment | | 231 | 28.10656 |
| M3-8-Best | Fine Tune Augment | | 54 | |
| M3-9 | Fine Tune Augment | 0.3 0.3 | 964 | 110.07629 |
| M3-9-Best | Fine Tune Augment | 0.3 0.3 | 766 | |
| M3-10 | Fine Tune Augment | 0.5 0.5 | 727 | 79.46 |
| M3-10-Best | Fine Tune Augment | 0.5 0.5 | 606 | |
| M3-11 | Fine Tune Augment | | 504 | 56.86579 |
| M3-11-Best | Fine Tune Augment | | 237 | |

As observed to the trained result in Table 1 elapsed minute is the time consumed from start to finish training. The fastest training time is the simple CNN with 4 layers. On the fourth and fifth column different value of dropout and epoch were used respectively to the models. VGG19 model consist of a large neural network with many parameters that's why it has the longest training time of 110.07629 minutes. The named "Best" in column 1 used early stopping and checkpoint model.

**Table 2. Accuracy and Training Loss**

| | Epoch | Loss | Acc | Val Loss | Val Acc |
|---|---|---|---|---|---|
| M1-1 | 208 | 5.83E-07 | 1 | 1.0059 | 0.8525 |
| M1-1-Best | 15 | 0.0898 | 0.968 | 0.4266 | 0.9098 |
| M1-2 | 210 | 3.98E-07 | 1 | 0.8855 | 0.918 |
| M1-2-Best | 32 | 0.0135 | 0.9959 | 0.4515 | 0.9344 |
| M1-3 | 213 | 0.0269 | 0.9948 | 0.8251 | 0.9098 |
| M1-3-Best | 191 | 0.007 | 0.9959 | 0.6332 | 0.959 |
| M1-4 | 255 | 2.34E-06 | 1 | 0.5055 | 0.9508 |
| M1-4-Best | 113 | 0.00002756 | 1 | 0.3045 | 0.9754 |
| M1-5 | 247 | 0.1157 | 0.9576 | 0.5523 | 0.9098 |
| M1-5-Best | 33 | 0.0151 | 0.9959 | 0.3278 | 0.9508 |
| M2-1 | 741 | 0.0503 | 0.9854 | 0.0169 | 0.9908 |
| M2-1Best | 401 | 0.0411 | 0.9896 | 0.0111 | 1 |
| M2-2 | 575 | 0.0436 | 0.9802 | 0.1321 | 0.968 |
| M2-2-Best | 375 | 0.0493 | 0.9875 | 0.0127 | 1 |
| M2-3 | 695 | 0.0526 | 0.9823 | 0.0903 | 0.9828 |
| M2-3-Best | 332 | 0.0676 | 0.9771 | 0.0211 | 1 |
| M3-4 | 318 | 0.2689 | 0.8686 | 0.1475 | 0.9462 |
| M3-4-Best | 76 | 0.2661 | 0.8874 | 0.1612 | 0.9680 |
| M3-5 | 728 | 0.1069 | 0.9562 | 0.1222 | 0.9591 |
| M3-5-Best | 440 | 0.1074 | 0.9593 | 0.0329 | 1 |
| M3-6 | 588 | 0.2826 | 0.876 | 0.157 | 0.9416 |
| M3-6-Best | 311 | 0.2776 | 0.8728 | 0.1634 | 0.9622 |
| M3-7 | 662 | 0.089 | 0.9649 | 0.1163 | 0.9771 |
| M3-7-Best | 233 | 0.1707 | 0.9322 | 0.0827 | 0.9920 |
| M3-9 | 964 | 0.1577 | 0.9371 | 0.1347 | 0.9577 |
| M3-9-Best | 766 | 0.2034 | 0.918 | 0.1048 | 0.9748 |
| M3-10 | 727 | 0.2358 | 0.8978 | 0.1682 | 0.9256 |
| M3-10-Best | 606 | 0.2831 | 0.8833 | 0.1598 | 0.9600 |
| M3-11 | 504 | 0.0862 | 0.9666 | 0.1514 | 0.9755 |
| M3-11-Best | 237 | 0.1389 | 0.9479 | 0.1258 | 0.9920 |

Overfitting is noticed when the training loss goes up meaning if the training loss got high the lower the training process. Malignant study[27] using CNN deep learning with the model Alexnet, VGG and with Area Under receiving Operating Characteristics (AUC) of 0.86 did not discuss overfitting and dropout. Table 2 uses different values of dropout as applied to the previous study that dropout at the max pooling layer[13] came up with a good result to overcome training loss. The resulted detection of abnormalities, such as polyps, ulcers[28]yet over-fitting still is not mentioned. Also in a polyp detection using deep learning model that reached 96%[17].

Table 2 shows that the highest accuracy percentage of 98.23% with less training loss of 9.03%. The observed best model is the Augmented (M2-3) with 0.5 0.5 dropout since accuracy is high while train loss is low. In the study that uses data augmentation controlled random noise to the training data[25]has also improve the CNN model[11] and each image is transformed into different position. Transfer of learning takes place when applying the previous model to another task during the generation of data or in feature extraction like in data augmentation. This concludes that applying augmented - pre-processing tasks and data augmentation techniques in deep learning offers better performance and outperforms other popular image processing techniques[11].

## 3.3 Optimization

Optimization in a deep neural network could improve performance. Adam is the optimization technique used in training all the models. It was observed that optimization to reduce only the style and content losses led to highly pixelated and noisy outputs[29].

Another way of improving the accuracy of the results is by tuning the machine leaning algorithm. Along the process of tuning dropout parameters was changed as resulted to different evaluated output and identified the best performance among the model.

Moreover, using VGG19 model many parameters are used compare from the simple CNN and Augmented model using a simple pre-trained model. VGG is one of the methods used in Transfer Learning. Like in the research works that incorporated popular Deep Learning (DL) architectures took advantage of transfer learning and this increases the learning efficiency of the problem by fine-tuning pre-trained models[25] like VGG16.

Different study applied the Transfer Learning with ResNet-50 for Malaria Cell-Image Classification[30]using fine-tune the pre-trained network (VGG19). It uses AlexNet and VGG 19 CNN architecture in SVM classifier, dropout, softmax, Relu, max pooling but again it did not discuss about overfitting, and evaluation accuracy. The mentioned study is different from this paper since Relu was activated and Adam was used all the time, sequential model uploaded from Keras and dropout was used interchangeably.

## 3.4 Performance Measures

The performance measures used are Confusion Matrix, Area Under the Curve – Receiver Operating Characteristics (AUC - ROC), and precision. AUC-ROC is the mostly used evaluation results for comparing different classification while confusion matrix is one of the performance measurements for machine learning classification[11].

**Table 3. Confusion matrix**

| Model | TN | FP | FN | TP | Accuracy (TP+TN) / (TOTAL) | Misclassification Rate (FP+FN) / TOTAL |
|---|---|---|---|---|---|---|
| M1-1 | 54 | 7 | 8 | 53 | 87.7% | 12% |
| M1-2 | 54 | 7 | 8 | 5 | 79.7% | 20% |
| M1-2-Best | 52 | 9 | 9 | 52 | 85.2% | 15% |
| M1-3 | 51 | 10 | 8 | 53 | 85.2% | 15% |
| M1-3-Best | 54 | 7 | 12 | 49 | 84.4% | 16% |
| M1-4 | 55 | 6 | 3 | 58 | 92.6% | 7% |
| M1-4-Best | 57 | 4 | 4 | 57 | 93.4% | 7% |
| M1-5 | 54 | 7 | 9 | 52 | 86.9% | 13% |
| M1-5-Best | 55 | 6 | 6 | 55 | 90.2% | 10% |
| M2-1 | 58 | 3 | 6 | 55 | 92.6% | 7% |
| M2-Best | 61 | 0 | 4 | 57 | 96.7% | 3% |
| M2-2 | 58 | 3 | 0 | 61 | 97.5% | 2% |
| M2-2-Best | 60 | 1 | 1 | 60 | 98.4% | 2% |
| M2-3 | 61 | 0 | 0 | 61 | 100.0% | 0% |
| M2-3-Best | 60 | 1 | 1 | 60 | 98.4% | 2% |
| M3-4 | 57 | 4 | 7 | 54 | 91.0% | 9% |
| M3-4-Best | 57 | 4 | 7 | 54 | 91.0% | 9% |
| M3-5 | 60 | 1 | 8 | 53 | 92.6% | 7% |
| M3-5-Best | 60 | 1 | 8 | 53 | 92.6% | 7% |
| M3-6 | 60 | 1 | 11 | 50 | 90.2% | 10% |
| M3-6-Best | 59 | 2 | 10 | 51 | 90.2% | 10% |
| M3-7 | 59 | 2 | 5 | 56 | 94.3% | 6% |
| M3-7-Best | 59 | 2 | 6 | 55 | 93.4% | 7% |
| M3-9 | 60 | 1 | 10 | 51 | 91.0% | 9% |
| M3-9-Best | 56 | 5 | 5 | 56 | 91.8% | 8% |
| M3-10 | 59 | 2 | 10 | 51 | 90.2% | 10% |
| M3-10-Best | 58 | 3 | 10 | 51 | 89.3% | 11% |
| M3-11 | 60 | 1 | 6 | 55 | 94.3% | 6% |
| M3-11-Best | 59 | 2 | 9 | 52 | 91.0% | 9% |

Accuracy in Table 3 means it is the prediction identified correct and incorrect of the True Positive (TP) add to True Negative (TN) all over Total. Still as shown in Table 3 M2-3 has 0 misclassification rate among models.

Table 4 Shows the Sensitivity, Precision, and Specificity of the models, F1 Score and Receiver Operating Characteristics of each model.

**Table 4. Sensitivity, Precision, Specificity, F1 Score & Receiver Operating Characteristics of the Models**

| Model | Sensitivity / Recall TP / Actual Yes | Precision TP / (TP + FP) | Specificity TN / Actual No (%) | F1 Score (%) | ROC % |
|---|---|---|---|---|---|
| M1-1 | 87 | 89 | 89 | 88 | 93 |
| M1-1-Best | 75 | 85 | 85 | 80 | 88 |
| M1-2 | 38 | 89 | 89 | 88 | 92 |
| M1-2-Best | 85 | 85 | 85 | 85 | 91 |
| M1-3 | 87 | 84 | 84 | 85 | 89 |
| M1-3-Best | 80 | 89 | 89 | 84 | 92 |
| M1-4 | 95 | 90 | 90 | 93 | 97 |
| M1-4-Best | 93 | 93 | 93 | 93 | 97 |
| M1-5 | 85 | 89 | 89 | 87 | 92 |
| M1-5-Best | 90 | 90 | 90 | 90 | 95 |
| M2-1 | 90 | 95 | 95 | 93 | 99 |
| M2-Best | 93 | 100 | 100 | 97 | 100 |
| M2-2 | 100 | 95 | 95 | 98 | 100 |
| M2-2-Best | 98 | 98 | 98 | 98 | 100 |
| M2-3 | 100 | 100 | 100 | 100 | 100 |
| M2-3-Best | 98 | 98 | 98 | 98 | 1 |
| M3-4 | 89 | 93 | 93 | 91 | 97 |
| M3-4-Best | 89 | 93 | 93 | 91 | 96 |
| M3-5 | 87 | 98 | 98 | 93 | 98 |
| M3-5-Best | 87 | 98 | 98 | 93 | 98 |
| M3-6 | 82 | 98 | 98 | 90 | 97 |
| M3-6-Best | 84 | 97 | 97 | 90 | 96 |
| M3-7 | 92 | 97 | 97 | 94 | 99 |
| M3-7-Best | 90 | 97 | 97 | 93 | 98 |
| M3-9 | 84 | 98 | 98 | 91 | 98 |
| M3-9-Best | 92 | 92 | 92 | 92 | 98 |
| M3-10 | 84 | 97 | 97 | 90 | 97 |
| M3-10-Best | 84 | 95 | 95 | 89 | 97 |
| M3-11 | 90 | 98 | 98 | 94 | 99 |
| M3-11-Best | 85 | 97 | 97 | 91 | 98 |

The precision in Table 4 identify the correctness of the prediction which is equal to TP / (TP+FP) while the specificity[31]means the proportion of the true negative occupying the actual negative. M2-3 model has the highest correctness rate of 100%.

In Table 2, the lowest accuracy percentage is 96.8% which is the Simple CNN (M1-1 Best) while the ROC in Table 4 reached 88%. M2-3 model got the highest ROC evaluation model of 100%.

## 4  RESULTS AND DISCUSSION

Figure 4 shows the simple CNN with high training accuracy and low validation accuracy in Figure 4(A) and 4(B) show low accuracy loss and high validation loss which indicates overfitting.

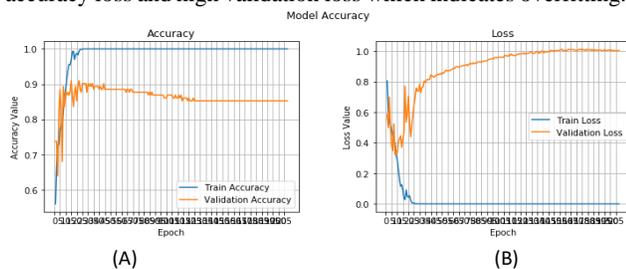

**Figure 4. Performance plot Simple CNN training overfits validation data.**

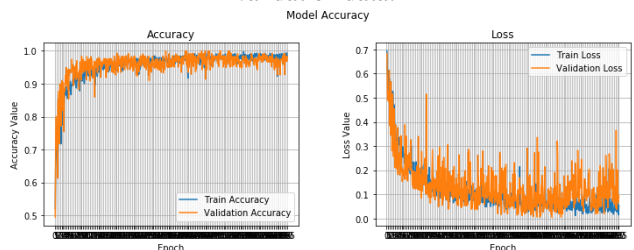

**Figure 5. Performance Plot with image augmentation and dropout 0.5 0.5**

The training loss was minimized in Fig. 5(B) because it plugged down to zero while the accuracy percentage goes up to 1. These results show the high performance for Augmenting data with dropout equivalent to 0.5 0.5.

The confusion matrix in Figure 6 shows 100 percent of predication model for polyp and non-polyp image while 100% ROC in Figure 7.

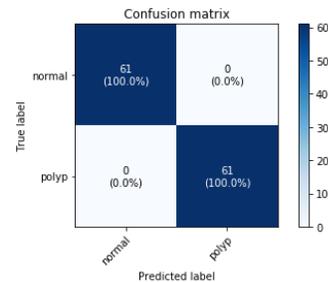

**Figure 6. Confusion Matrix of Augmented with Dropout 0.5**

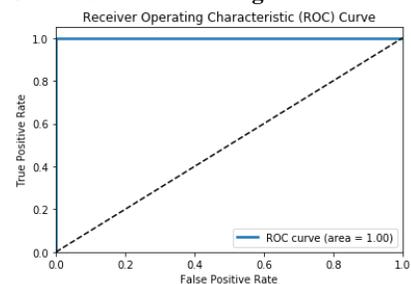

**Figure 7. Receiver Operating Characteristics (ROC) Curve**

Using a simple CNN layers with Adam and the different value of dropout to the series of model experiments came up with different results. This did adopt transfer learning as used by different study in detecting polyp and non-polyp image. Although previous study said that VGG19 is a more complex model as best-performing architectures[25] it is not the best resulted model but it is one of the three popular models such as Google Net and Residual Network. Simple 2CNN, Augmentation and VGG19 are the models used to randomly check which among them give the best performance based on the test results. It is concluded that the highest accuracy is the Augmentation with 0.5 dropout with the accuracy of 98.23% and in Figure 7 is the ROC curve of 100%. Transfer learning uses an adopted model (Simple CNN, Augmentation and VGG19) applied to another task as detecting polyp and non-polyp.

## 5  CONCLUSIONS

Technology advancement and application of transfer of learning in deep learning models have direct implications to medical detection and diagnostic like detecting polyp and non-polyp to images. Transfer of learning produces different high percentage of accuracy by experimenting different models in the CNN. Malaria Cell Image classification is one of the example that uses 160 million parameters with 95.91% accuracy[30].

Therefore, the study which states that it improves its results from exploring pooling layers, data augmentation, fully connected pre-trained CNN layer as a feature extractor without VGG and SVM[32] is correct since the highest accuracy from among the experiments made is the augmentation with dropout of 0.5 0.5. Transfer learning was used in this paper because it applies dropout, Adam optimizer, tuning model, and over fitting from previous studies.

It is said that the two main approaches to implement transfer learning are weight initialization like using pooling layer and the feature extraction[3] in VGG model.

Furthermore, different CNN model and algorithm can be used in application of transfer learning.

## ACKNOWLEDGMENT

I would like to acknowledge Engr. Alonica Villanueva for her unending suggestions and brainstorming sessions every time we meet; and to Engr. Menchie Miranda for her diligence and guidance in editing this paper.

**Arnel C. Fajardo, PhD** is a professor at the Computer Engineering Department of the Technological Institute of the Philippines in Quezon City. He finished his Doctor of Philosophy in Computer Engineering at Hanbat National University in 2014, Master's Degree in Computer Science at Dela Salle University in 1999 and Bachelor of Science in Electrical Engineering at the Mapúa Institute of Technology in 1991. His research interests include Artificial Intelligence, Image processing, Speech recognition, Software Engineering and Engineering Education. He has published more than 75 technical papers in Conferences and Journals

**Alexander A. Hernandez, DIT** is the Program Chair of the College of Information Technology Education at the Technological Institute of the Philippines in Manila. He holds a Doctor in Information Technology from the De La Salle University. His research interests include data mining, enterprise and decision support systems, sustainability and green computing. He has published papers in academic journals and conferences, including the International Journal of Enterprise Information Systems, Journal of Case on Information Technology, International Journal of Social Ecology and Sustainable Development, International Journal of Green Computing, International Journal of Socio-technology and Knowledge Development, among others.

**Ariel E. Isidro** is an Industry Lecturer at the Computer Engineering Department of Technological Institute of the Philippines in Quezon City. He is currently taking up Doctor in Information Technology at the same Institute. He finished his Master of Technology Management at the University of the Philippines in Diliman in 2004 and Bachelor of Science in Computer Engineering at the Mapúa Institute of Technology in 1987. His research interests include Computer Vision, Image Processing, Software Engineering, Systems Engineering and Engineering Education